\documentclass[%
preprint,
 amsmath,amssymb,
 aps,
]{revtex4-2}

\usepackage{graphicx}
\usepackage{dcolumn}
\usepackage{bm}

\begin{document}

\preprint{ARK-BECNondegNLDimer}

\title{Quantum Oscillations between weakly coupled Bose-Einstein Condensates:  Evolution in a Non-degenerate Double Well}

\author{John D. Andersen}%
 \email{jdasps@rit.edu}
\affiliation{School of Physics and Astronomy, 
Rochester Institute of Technology,
Rochester, New York   14623
}%
\author{Srikanth Raghavan}
\email{srikanth.raghavan@navy.mil; S.R. contributed to this article in his personal capacity. The views expressed are his own and do not necessarily represent the views of the United States, the Department of Defense (DOD), or its components.}
\affiliation{Naval Surface Warfare Center, Carderock Division, West Bethesda, MD 20817}
\author{V. M. Kenkre}
\email{kenkre@unm.edu}
\affiliation{Department of Physics and Astronomy, University of New Mexico, Albuquerque, NM 87131}




\date{\today}

\begin{abstract}
We discuss  coherent atomic oscillations between two weakly coupled Bose-Einstein condensates that are energetically different. The weak link is notionally provided by a laser barrier in a (possibly asymmetric) multi-well trap or by Raman coupling 
between condensates in different hyperfine levels. The resultant boson Josephson junction dynamics is described by a double-well nonlinear Gross-Pitaevskii equation. On the basis of a new set of Jacobian elliptic function solutions, we describe the period of the oscillations as well as associated quantities and predict novel observable consequences of the interplay of the energy difference and initial phase difference between the two condensate populations. 
\end{abstract}

\maketitle


\section{\label{sec:intro} Introduction}
As has been detailed in a variety of papers, \cite{sfgs1997,rsfs1999,mcww1997,rsk1999}, a bosonic Josephson junction (BJJ) is a device in which two or more macroscopic ensembles of bosons \cite{rsfs1999}, each of them occupying a single quantum state, are coupled to a tunnel barrier.
As is well-known, the order parameter identifying the one-body macrosopic condensate wave function, $\Psi(r,t)$, obeys a nonlinear Schr{\"o}dinger equation known in the literature as the Gross-Pitaevskii equation (GPE) \cite{gpepitaevskii1961,gpegross1961,dspsrmp1999,leggettrmp1999}
\begin{equation}
\label{eq:GPEbasic}
i\hbar\frac{\partial \Psi(r,t)}{\partial t} = -\frac{\hbar^2}{2 m} \nabla^2 \Psi(r,t) + \left[V_{ext} + g_0 \left|\Psi(r,t)\right|^2\right] \Psi(r,t),
\end{equation}
with $m$ the atomic mass, $V_{ext}$ is the external potential, $g_0 = \frac{4\pi\hbar^2 a}{m}$, and $a$ the $s$-wave scattering length of the atoms.  We consider in the following, specifically a double-well (state) system where the barrier is created by a far-off-resonance laser barrier that cleaves a single trapped condensate into two possibly asymmetric parts \cite{andrews1997,albiez2005,llss2007}, and report on some interesting predictions that we make on the basis of recently developed analytic methods on a well-known system of long standing interest. The results could well apply to other condensate systems as well.     
\par
Following the analysis of  \cite{rsfs1999,sfgs1997}, we write a (time-dependent) variational wave function for the total condensate as 
\begin{equation}
\label{eq:2stateBECdefn}
\Psi(r,t) = \psi_1(t) \Phi_1(r) +\psi_2(t) \Phi_2(r)  ,
\end{equation}
where
\begin{eqnarray}
\label{eq:psi12defn}
\psi_i(t) &=& \sqrt{N_i(t)} e^{i\theta_i (t)}, \; \; i=1,2\\
\sum_{i=1}^{2} N_i &=& \sum_{i=1}^{2} |\psi_i|^2 \equiv N_T,
\end{eqnarray}
$N_T$ being the constant total number of condensate atoms. The amplitudes for the individual condensate populations obey the two-mode dynamical equations \cite{rsfs1999,sfgs1997,mcww1997,zaptasolsleggett1998,rsk1999,kc1986}
\begin{subequations}
\label{eq:NonDegenAmplitudeAllequations} 
\begin{eqnarray}
i\hbar \frac{\partial \psi_1}{\partial t} &=& \left(E_1^0 + U_1 N_1\right) \psi_1 + {\mathcal K}\psi_2, \label{eq:NonDegenDimerequationa}
\\
i\hbar \frac{\partial \psi_2}{\partial t} &=& \left(E_2^0 + U_2 N_2\right) \psi_2 + {\mathcal K}\psi_1, \label{eq:NonDegenDimerequationb}
\end{eqnarray}
\end{subequations}
In Eqs~(\ref{eq:NonDegenAmplitudeAllequations}), $E_i^0$ are the ground-state energies in each well (state), $U_{i} N_{i}$ are proportional to the atomic self-interaction energies and ${\mathcal K}$ describes the tunneling amplitude between condensate states. These constant parameters can be written in terms of the the spatial wavefunctions $\Phi_i(r)$ as
\begin{subequations}
\label{eq:NonDegenParametersAllequations} 
\begin{eqnarray}
E_i^0 &=& \int \frac{\hbar^2}{2m} |\nabla \Phi_i|^2 + |\Phi_i|^2 V_{ext}(r) dr ,\label{eq:NonDegenParametersequationa}
\\
U_i &=& g_0 \int  |\Phi_i|^4 dr,\label{eq:NonDegenParametersequationb}
\\
{\mathcal K}  &=& \int \left[ \frac{\hbar^2}{2m} \left( \nabla \Phi_1 \cdot \nabla \Phi_2 \right) + \Phi_1 V_{ext} \Phi_2  \right]  dr.\label{eq:NonDegenParametersequationc}
\end{eqnarray}
\end{subequations}
If we assume that the the single trapped condensate is split into two asymmetric parts \cite{andrews1997} but allow for the dynamics to be such that the population difference between the two condensates not to significantly affect the spatial structure, to lowest order, we may assume that
the spatial parts of the wavefunction $\Phi_i(r)$ do not change appreciably  \cite{rsfs1999,sfgs1997}. Further, by re-scaling time, $t \rightarrow t/\hbar$, and making the following definitions for the inter-well (state) coupling constant, nonlinear, energy parameters, and 
the well amplitudes,
\begin{subequations}
\label{eq:Delta-Chi-V definition}
\begin{eqnarray}
V &\equiv& \frac{{\mathcal K}}{\hbar}, \\
\chi &\equiv& -\frac{(U_1 +  U_2)N_T}{2 \hbar},\\
\Delta &\equiv& \frac{ 2(E_1^0 - E_2^0) + (U_1 -  U_2)N_T}{2 \hbar},\\
c_i &=& \psi_i/\sqrt{N_T}, i=1,2
\end{eqnarray}
\end{subequations}
and the usual Bloch vector definitions
\begin{equation}
p =|c_1|^2 - |c_2|^2, \; q = i(c_1^\ast c_2 - c_1 c_2^\ast), \;
r = c_1^\ast c_2 + c_1 c_2^\ast \label{eq:pqrdef}
\end{equation}
we obtain the obvious evolution equations for a nondegenerate double-well system,
\begin{subequations}
\label{eq:NondegDimerDNLSEAllequations_pqr} 
\begin{eqnarray}
\dot{p} &=& 2 V q,\label{eq:NondegDimerDNLSEpqrequationa}
\\
\dot{q} &=& -2 V p + (\Delta - \chi p) r ,\label{eq:NondegDimerDNLSEpqrequationb}
\\
\dot{r} &=& -(\Delta - \chi p) q.\label{eq:NondegDimerDNLSEpqrequationc}
\end{eqnarray}
\label{bloch}
\end{subequations}
Equations~(\ref{eq:NondegDimerDNLSEAllequations_pqr}) describe at least three possible Bose-Einstein condensate systems. One could consist of condensates created from cold atoms (e.g. Na, Rb) in the same internal state but confined in a double-well trap produced, for example, by a far off-resonance laser barrier that cuts a single trapped condensate into two asymmetric parts \cite{andrews1997,AlbiezPhD2005}. A different possible realization could be a two-component system of condensates, e.g. from the $|F=1, m_f = -1\rangle$ and $|2,1\rangle$ spin states of $^{87}\mbox{Rb}$ \cite{hall1998,williams1999,rsfs1999,smerziEPJB2003}. Yet a third realization might be by polariton condensates confined in a photonic molecule formed by two overlapping micropillars etched in a semiconductor microcavity \cite{galbiati2012,scws2008,sspm2008,abbarchi2013} wherein the static energy difference $\Delta E$ is the
energy difference between the ground states of the left and right micropillars in the absence of coupling. 

The purpose of this paper is to present analytical results and explicit plots we have produced with particular focus on observable quantities in these systems. They appear below as four items: the study of statics resulting in stationary states of the system that is described in Section II; and the study of dynamics presented in Section III directed at three observables dealing, respectively, with the dependence of the frequency of quantum oscillations on nonlinearity, the movement of critical points and a critical line of transitions in parameter space, and the time dependence, in some cases peculiar, of the population difference between the two wells (states) of the condensates. Our hope is that experimentalists will use these predictions we make on the basis of our theoretical analysis  to compare with their observations and thereby probe the domain of validity of the belief current in the literature in the applicability of the underlying theoretical concepts.

\section{\label{sec:statstates} Statics of the Nondegenerate Boson Josephson Junction: Stationary States and their Stability}
States of the system under study that have no time dependence whatsoever are called stationary states. If the system occupies any of them initially, it remains in them forever. Additionally, if we append a damping agency to the system that naturally removes energy or excitation from it, the expectation is that the system settles in these stationary states at long enough times. Also for these reasons, they are the states that are easily accessible to external probes. Finding these states and determining the extent of their stability are, consequently, important undertakings.

We arrive at the stationary states by putting all time derivatives in Eq. (\ref{bloch}) to zero. Such a procedure (see, e.g., \cite{kc1986,ktc1987}) is equivalent to the method followed by the first investigators of stationary states of the discrete nonlinear Schr\"{o}dinger equation (\cite{els1985}) and produces from (\ref{bloch}a) and (\ref{bloch}c) the simple consequence that $q_{ss}=0$: we denote stationary state values of the Bloch components by the subscript $ss$. On the other hand, Eq.~(\ref{bloch}b) yields, on equating the time derivative in it with zero, an interesting relation between the value $p_{ss}$ of the probability difference and the parameters of the system. These parameters are the inter-well (state) coupling constant $V$, the static energy difference $\Delta$ and the nonlinearity $\chi$. We find the relation that delineates the stationary states to be
\begin{equation}
\frac{\chi}{2V}=\frac{\delta}{p_{ss}}-\frac{1}{r_{ss}}=\frac{\delta}{p_{ss}} \mp\frac{1}{\sqrt{1-p^2_{ss}}}.
\label{ssnond}
\end{equation}
In determining it, we make use of $q_{ss}=0$ and of the conservation of probability that results in $p^2_{ss}+q^2_{ss}+r^2_{ss}=1$, and introduce the notation  $\delta=\Delta/2V$.

Stability of the stationary states is studied by expanding the three components of the Bloch vector around those stationary states as $$p=p_{ss}+\epsilon_p,\,\,\,\,\,\,\,\,\,\,\,\,q=q_{ss}+\epsilon_q=0+\epsilon_q,\,\,\,\,\,\,\,\, \,\,\,\,  r=r_{ss}+\epsilon_r,$$ 
and linearizing the equations of motion in the variations $\epsilon_{p,q,r}$. By finding the eigenvalues of the relevant relaxation matrix, 

\begin{equation}
\frac{d}{d\tau}
\begin{pmatrix}
  \epsilon_p \\
 \epsilon_q\\
  \epsilon_r
\end{pmatrix}
=
\begin{pmatrix}
  0 & 1 & 0\\
  -(1+2k_0r_{ss}) & 0 & +\frac{p_{ss}}{r_{ss}} \\
   0 & -\frac{p_{ss}}{r_{ss}} & 0. 
\end{pmatrix}
\begin{pmatrix}
  \epsilon_p \\
  \epsilon_q \\
  \epsilon_r 
\end{pmatrix},
\end{equation}
we determine its stability. If any of the three eigenvalues of the matrix has a positive real part, the stationary state is unstable. The eigenvalues $\lambda$ are, explicitly,
\begin{equation}
-\lambda[\lambda^2+(p_{ss}/r_{ss})^2]-(1)[(1+2k_0r_{ss})\lambda-0]=0.
\end{equation}
One of the eigenvalues is zero. The other two are 
\begin{equation}
\lambda^2=-(2k_0p_{ss}-\delta)^2+\frac{\delta}{2k_0p_{ss}-\delta},
\end{equation}
where we have introduced the dimensionless ratio $k_0=\chi/4V$. This then allows us to conclude that the stationary state is unstable if 
\begin{subequations}
\begin{align}
\delta/2k_0 < p_{ss}<(\delta^{1/3}+\delta)/2k_0,\,\,\,\mbox{for}\,\,\delta >0,\\
\delta/2k_0 > p_{ss}>(\delta^{1/3}+\delta)/2k_0,\,\,\,\mbox{for}\,\,\delta <0.
\end{align}
\label{wufab}
\end{subequations}
We have found that such an analysis was originally carried out by Wu \cite{WuPhD1989} and  we believe that it has not been published in the open literature. 
\begin{figure}[]
\begin{center}
\includegraphics[width=0.6\textwidth]{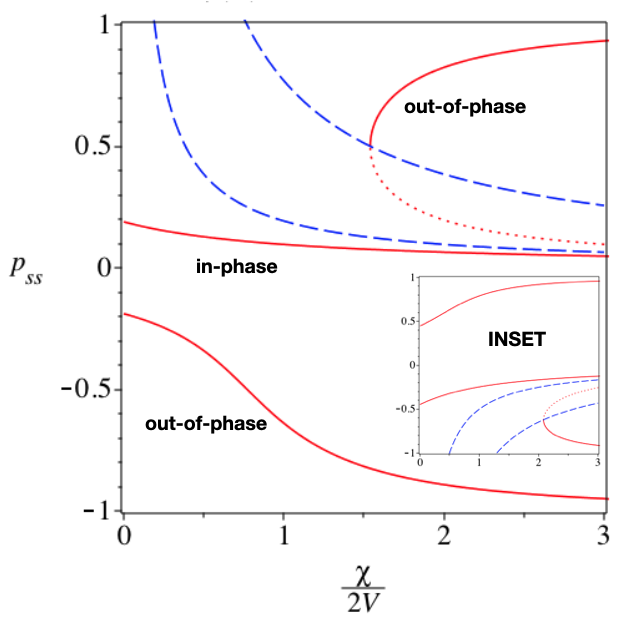}
\caption{Stationary state probability $p_{ss}$ of the system (solid red line) obtained as a function of the nonlinearity ratio $\chi/2V$ obtained by putting the time derivatives in Eqs. (\ref{bloch}) equal to zero. The main (inset) plot is for a positive value $1/\sqrt{27}$ (negative value $-1/2$) of the static energy mismatch relative to $2V$. The inset is much like the main plot except that it is largely inverted vertically with respect to the latter. Dotted lines represented unstable stationary states and the dashed (blue) lines contain the parameter space in which the states are unstable. See text including Eqs.~(\ref{wufab}).}
\label{fig:statfig}
\end{center}
\end{figure}

We display the results here in Fig. \ref{fig:statfig} where the probability difference in the stationary states is plotted as a function of $\chi/2V$ for two values of the static energy mismatch: a positive value $\delta=\Delta/2V=1/\sqrt{27}$ in the main plot and a negative value $\delta=-1/2$ in the inset (for the same axes and scales as in the main plot). Given that the inset shows only slight changes of values and an up-down inversion because of the change of sign, let us focus our discussion only on the main plot. The solid lines represent  the stable stationary states while the dotted lines represent the unstable situation. The two lines marked `out-of-phase' have $r_{ss}=-\sqrt{1-p^2_{ss}}$ while for those marked `in-phase',  $r_{ss}=+\sqrt{1-p^2_{ss}}$. The dashed lines are an unusual feature of the stationary state analysis: they set the limits that mark the region in which the stationary states are unstable as dictated by Eqs. (\ref{wufab}). 

We hope that a well-thought out probing of the dependence of the stationary state probabilities $p_{ss}$ on the nonlinearity ratio might be observationally possible with the help of Fig. \ref{fig:statfig} as a guide. It might be possible to observe the time-dependent relaxation of the coherent oscillations of the system into these stationary states as well as excitation of the system out of the stationary states with the help of an external probe.  In a different system obeying formally similar evolution equations
\cite{KenWuPhysLett1989,*KenWuPRB1989}, initial oscillations characteristic of the undamped system and eventual settling into the stationary states characteristic of the damped system have shown some remarkably interesting behavior.

\section{\label{sec:nonlinearBJJdynamics} Dynamics:  Frequency Variation,  Movement of Critical Points and Time-dependence}
While the potential method of solution of Eqs.~(\ref{bloch}) typically involves Weierstrass elliptic functions \cite{TsiPhysLett1993}, we have recently found equivalent Jacobian elliptic forms that are considerably more convenient. The enhanced convenience will become apparent when we describe some of the characteristics of the behavior. The general solution for  the fractional population difference between the two condensate states $p(t)=P_1(t)-P_2(t)$ is, with the notation $\tau=2Vt$,
\begin{equation}
\frac{p(\tau)}{p(0)}=1-B\left[1-\frac{1}{1+a\,\mbox{sn}^2\left(\tau\sqrt{e_1-e_3},k_1\right)}\right],
\label{ptauje}
\end{equation}
where $p(0)$ denotes the initial fractional population difference, $\mbox{sn}(u,k)$ denotes the Jacobian elliptic sine function with the elliptic \textit{modulus} $k$ (or equivalently the elliptic \textit{parameter} $m=k^2$) \cite{byrd-friedman}. The quantities $e_i$ (with $i=1, 2, 3$), as also $a$ and $B$, can be obtained easily from the system parameters and initial conditions. See below
\footnote{The quantities $e_i$ are roots of the equation
\[4\sigma^3-g_2\sigma-g_3=4(\sigma-e_1)(\sigma-e_2)(\sigma-e_3)=0,\]
and determine the elliptic parameter $k_1$ via
\[k_1^2=\frac{e_2-e_3}{e_1-e_3}.\]
The quantities $g_1$ and $g_2$  are Weierstrass entities defined in terms of the system parameters and of the value $U_0$ of the potential $U(p)$ at $p=p(0)$ as 
\[g_2=3\gamma_1^2+2\gamma_0\delta k_0-k_0^2U_0,\,\,\,\,\,\,\,\,\,\,
g_3=\gamma_1^3+\gamma_0^2k_0^2+\gamma_0\gamma_1\delta k_0+(\gamma_1k_0^2-\delta^2 k_0^2/4)U_0.\]
The quartic potential itself is
\[U(p)=
 4\gamma_0p
+6\gamma_1p^2
-2\delta k_0p^3
+k_0^2p^4.\]
The coefficients $\gamma$ are obtained from system parameters:
\[\gamma_0=\delta\left(p(0)^2k_0-p(0)\delta-r(0)\right)/2
),\,\,\,\,\,\,\,\,\,\,\gamma_1=\left(-2p(0)^2k_0^2+2k_0r(0)+1+2 p(0)\delta k_0+\delta^2\right)/6.\]
Finally, $a$ and $B$ are obtained from
\[a=\frac{24e_3+U''_0}{24(e_1-e_3)},\,\,\,\,\,\,\,\,\,\,B=\frac{6U_0'}{p(0)(24e_3+U_0'')},\]
 where $r(0)$ denotes the initial value of $r(t)$, $U_0'$ and $U''_0$ are the first and second $p$-derivatives of the quartic potential  $U(p)$ evaluated at $p=p(0)$.}.

Equation (\ref{ptauje}) allows us to  calculate, easily \textit{and analytically}, both the period $T$ of $p(\tau)$ and (especially)  its time-averaged  value over a cycle, $\langle p(\tau)\rangle$. We obtain the latter as 
\begin{equation}
\langle p(\tau)\rangle=p(0)\left[1-B\left(1-\frac{\Pi(-a;\alpha,k_1)}{\alpha}\right)\right]
\label{pavell}
\end{equation}
in terms of the incomplete elliptic integral of the third kind defined through
$$\Pi(n;u,k)=\int_0^u \frac{1}{1-n\mathrm{sn}^2(w,k)}dw.$$
The quantity $\alpha$ is related to the period through  $\alpha=T\sqrt{e_1-e_3}/2,$ and the period $T$ is given by
\begin{equation}
T=\sqrt{2}(12e_1^2-g_2)^{-1/4}\mathrm K(k_2)
\label{periodell}
\end{equation}
in terms of the complete elliptical integral of the first kind, $K,$ with its argument $k_2$ given by
\begin{equation}
k_2^2=\frac12\left(1-\frac{3e_1}{(12e_1^2-g_2)^{1/2}}\right).
\label{k2sq}
\end{equation}
In Eq.~(\ref{k2sq}), $e_1$ is the largest real root of the three $e_i$. The average value given by Eq. (\ref{pavell}) is useful in delineating special values in parameter space. The analytic expression (\ref{periodell}) for the period is used directly in the plots we show. 

These novel analytical results of our study allow us to investigate, and display graphically for the observer, three plots as shown below that make explicit the predicted variation of measurable quantities.

\subsection{Behavior of the Frequency of the Quantum Oscillations beyond the Critical Point}
\begin{figure}[h]
\begin{center}
\includegraphics[width=0.9\textwidth]{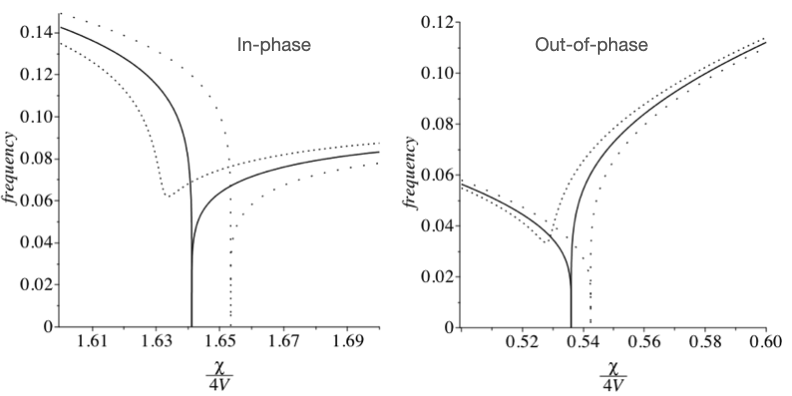}
\caption{Dependence of the frequency  of quantum oscillations  (essentially reciprocal of the period of $p(t)$) showing the disappearance of the selftrapping transition near the critical point as the static energy mismatch is made sufficiently negative. The in-phase (left panel) and out-of-phase (right panel) cases (see text) are qualitatively identical but differ slightly in shape and in location of the behavior change on the nonlinearity ($\chi/4V$) axis. The middle (solid) line in either case is at the critical point. The rightmost curves (space dotted) in either panel also show a selftrapping transition as do the middle lines. There is no transition in the leftmost curves (dotted). The location of the transition on the nonlinear axis is at higher (lower) value of the nonlinearity relative to the resonant dimer case (at $\chi/4V=1$, not shown). }
\label{fig:freq}
\end{center}
\end{figure}

In the case of a \textit{resonant} pair of states (i.e. when both states have the same energy), we understand easily that greater initial occupation of state 1 (relative to state 2) will initially make initially state 1 lower in energy than state 2 (assuming infinite relaxation to the lowered energy). For a sufficiently large nonlinearity, the two states will be unable to come into resonance dynamically as transfer proceeds. Consequently, a selftrapping transition will occur. This  scenario continues to hold for an asymmetric double well (state) condensate system for positive $\delta=\Delta/2V$: there is always the occurrence of self-trapping for all large enough $\chi$. The situation for negative $\delta$ is more subtle. It can be shown as a consequence of Eqs.~(\ref{bloch}) that selftrapping will occur at values of $\chi$ lower than that necessary for the symmetric double well (state) system, but that below a certain threshold, the possibility of a transition will disappear altogether. As one enters this region, one will not encounter singular behavior at a point in $\chi$-space in the average value  of $p(t)$ over a cycle or of its frequency. Instead, the change will smooth out and eventually, for sufficiently negative mismatch, all signs of the transition will disappear.
 
We have found in our study of the past literature that this remarkable evolution, marked by what may be called a `critical point' was first discovered by Tsironis \cite{TsiPhysLett1993,MolTsi1995}, and  can be obtained exactly from Eqs. (\ref{bloch}), specifically from the solution (\ref{ptauje}) that we have presented. This is shown in Fig. \ref{fig:freq}.  The initial probability difference we consider here is $p(0)=0.6.$ The selftrapping transition is clear in Fig. \ref{fig:freq} in both the left panel (in-phase) and the right panel (out-of-phase)  and is depicted by the right two curves in each panel. The left-most curve in each panel, however, shows that there is no selftrapping occurring: the curves have no singular behavior even though they change direction (slope changes sign) at the expected location of the change. Going from left to right in the left panel, the three curves are calculated for the energy mismatch ratio $\delta=-1.3274, -1.3287, -1.3301$ while in the right panel it is for $\delta=-0.0133, -0.0103, -0.0073$ in order. The middle curve in each case represents the Tsironis critical point.

The frequency here is the reciprocal of the quantity defined in Eq. (\ref{periodell}) and its analytical evaluation and plotting is facilitated by the new Jacobian elliptic function solutions we have provided here. We hope that this variation of the frequency with nonlinearity that the model predicts can be measured experimentally via a direct experimental probe, properly designed. 
\begin{figure}[h!]
\begin{center}
\includegraphics[width=0.6\textwidth]{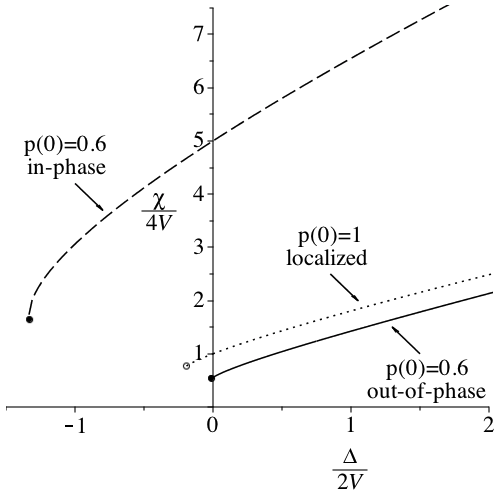}
\caption{Movement of the critical point and the entire critical line (depicting selftrapping transitions) which ends in the critical point as the initial occupation is changed. The middle (dotted) line in the Figure (ending in an open circle) is what we call the Tsironis line: each point on it marks a self-trapping transition as one moves over the plane made by the dynamic mismatch parameter $\chi/4V$ and the static counterpart $\Delta/2V$. We show how the critical point and the entire critical line move with initial occupation. This is particularly relevant to  Bose-Einstein condensate experiments because an initially localized condition is not attainable experimentally. See text.}
\label{fig:crit}
\end{center}
\end{figure}

We now come to our prediction for how the critical point, and indeed the entire transition line, moves in parameter space as we change the initial distribution $p(0)$ between the two sites.

\subsection{Movement of the Critical Point and the Critical Line in Parameter Space}
The self-trapping transition is marked by the time-dependence of $p$ changing from being oscillatory to a hyperbolic decay as the nonlinearity ratio is increased to the critical value  and then changing back to oscillatory as the nonlinearity goes beyond that value. Our analytic expression (\ref{periodell}) for the period of $p$ makes it easy to determine this passage. Figure~\ref{fig:crit} shows this happening. The dotted line represents an initially localized condition $p(0)=1$. It has little relevance to BEC observations because arranging for localized initial conditions is practically impossible. To be able to analyze a feasible situation, we have taken $p(0)=0.6$ throughout this paper wherever an initial condition has been used. For the sake of simplicity and analytic convenience, we take real conditions which means that $q(0)=0$. There are then two possibilities identified as $r(0)=\pm \sqrt{1-p(0)^2}$ since conservation of probability demands that the sum of the squares of the three components of the Bloch vector add up to 1.  We have already referred to these two cases as, respectively, the in-phase and out-of-phase cases.

In Fig. \ref{fig:crit}, the selftransition line for $p(0)$ \cite{TsiPhysLett1993,MolTsi1995} is shown dotted.  It obeys 
\begin{equation}
8\delta k_0^3-(12\delta^2-1)k_0^2+2\delta(3\delta^2-5)k_0-(\delta^2+1)^2=0
\label{georgeline}
\end{equation}
but, as explained above, its assumption of initial condition is not appropriate in the BEC context.  The results of our recent calculations for $p(0)=0.6$ are shown as a solid line for out-of phase and as a dashed line for in-phase initial conditions. These, by contrast, are accessible experimentally in BEC systems. From our general theoretical expressions  such as Eq. (\ref{ptauje}), we are able to derive their equations explicitly. With the understanding that the upper sign in the  $\pm$ or $\mp$ option stands for the in-phase and the lower sign for the out-of-phase case, they are given by
\begin{align}
\nonumber 576k_0^4
+40(3\delta\mp46)k_0^3
-25(92\delta^2\pm264\delta+183)k_0^2 \\ 
+250[9\delta^3
-15\delta\pm4(5\delta^2-3)]k_0
-625(\delta^2+1)^2=0
\end{align}

The movement of the critical line and the critical points (shown by circles at the end of the lines) as we change the quantum phase in the initial condition, should be clear from Fig. \ref{fig:crit}. It might be interesting to look for the characteristic time dependence of the critical line (hyperbolic secant except at the critical point where it reduces to power law form) when the system is on these lines. We are able to calculate the corresponding evolution and thus locate the transition for any given initial condition.

\subsection{Time-dependence of the Condensate Population Difference $p(\tau)$}
\begin{figure}[h!]
\begin{center}
\includegraphics[width=0.8\textwidth]{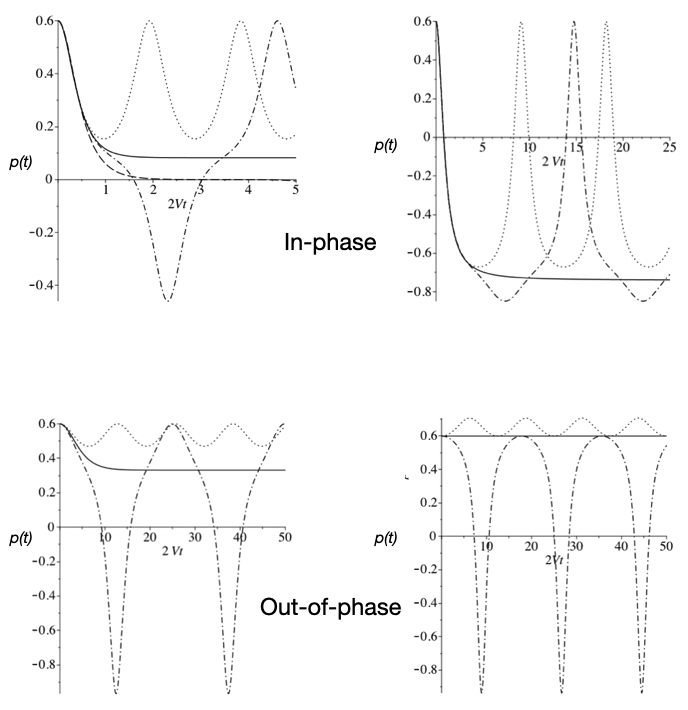}
\caption{Dependence of $p$ on the dimensionless time $\tau=2Vt$ from solutions of Eq. (\ref{bloch})  in situations where they have a markedly different appearance from those seen in ordinary situations for a degenerate nonlinear dimer. The dimensionless parameter values of $\Delta/2V$ and $\chi/4V$ for each panel, moving clockwise from the top left, and in sequential order by line type (dotted, solid, dash-dotted), are in the format $[\Delta/2V;\chi/4V$]  as follows: [1;(6.6666,6.552,6.526)],[-1.329;(1.654,1.641,1.633)],[27/64;(0.9863,125/128,0.9688)], and [1/10;(0.6877,0.6809,0.6741)].}
\label{fig:time}
\end{center}
\end{figure}

The solution  Eq.~(\ref{ptauje}), that we have obtained  in the Jacobian elliptic form, makes it easy for us to examine features of the time dependence of $p$. Figure~\ref{fig:time} displays them in special situations that show their marked difference from counterparts in the degenerate nonlinear dimer. In the upper panel we show a case of in-phase real initial conditions which means, as we have already explained, $q(0)=0, r(0)=+\sqrt{1-p(0)^2}$ while in the lower panel we have out-of-phase real initial conditions which means $q(0)=0, r(0)=-\sqrt{1-p(0)^2}$. In both panels we have taken $p(0)=0.6$. We have taken care to avoid the more easily studied localized initial conditions ($p(0)=1$) because it is practically impossible to realize them in a Bose-Einstein condensate experiment.

The solid lines throughout the four cases shown display the hyperbolic secant transition evolution of $p(\tau)$ at the point where self-trapping occurs. The dotted lines in each case show oscillations corresponding to nonlinearity that exceeds the transition value slightly while the dash-dotted lines show oscillations where the nonlinearity is slightly exceeded by its transition value. While the top left panel (displayed for comparison because of its similarity to the situation in the \textit{degenerate} double-well) shows unremarkable behavior, the variation in the other three panels is quite worthy of note because of strong changes (e.g., spikes in the curves) that would be interesting to observe in an experiment. 

The top two panels, which happen to have been based on calculations with in-phase initial conditions, should be viewed together to appreciate the distinct differences that nondegeneracy can introduce into the evolution of $p$. In the left panel it is quite similar to the degenerate case. To emphasize this similarity we have included in the figure the hyperbolic secant transition curve (dashed line) that represents self-trapping in the degenerate system. The other three curves do not differ markedly from  degenerate counterparts: the system oscillates equally on the two sides (dash-dotted line) if the nonlinearity is below the critical value and only one side if above it (dotted line). 

\textit{By contrast}, the evolution is quite different in appearance in the right panel. We are here at the critical point in the nondegenerate case. In the vertical direction the evolution is sharply different in that there are narrow spikes at the top and different shapes at the bottom. Such differences should be discernible experimentally by traveling the parameter space.

The lower two panels in Fig. \ref{fig:time} show another kind of striking behavior. They have been computed for out-of-phase initial conditions. These alone can produce an \textit{amplitude} transition in addition to the usual self-trapping transition. The possibility of an amplitude transition was shown long ago for polaronic nonlinear dimers \cite{TsiKen1988,*Ken1993} and more recently in Bose-Einstein condensate systems \cite{rsk1999} as representative of the initial state \textit{coinciding} with the stationary state of the system. This is possible only for out-of-phase initial conditions. While the two transitions generally occur at different values of the nonlinearity parameter, it is possible in the \textit{nonresonant} system to make the values identical by controlling $\Delta$, the static mismatch in energy. Such a situation, in which the critical values of $\chi$ for the two transitions  is identical, is particularly interesting and  is depicted in the right lower panel. Notice that the initial occupation of $p_0=0.6$ is seen to continue without \textit{any change} throughout the evolution because we have hit upon a stationary state. 

For the situation in which the two transitions coincide, our theory has a clear prediction for the relationship between the initial value $p(0)$ and the system parameters $\delta=\Delta/2V$ and $k_0=\chi/4V$ that describe, respectively, the static and the dynamic energy mismatch:
\begin{equation}
\left[\delta, k_0\right]=\left[\frac{p(0)^3}{(1-p(0)^2)^{3/2}},\,\frac{1}{2(1-p(0)^2)^{3/2}}\right].
\end{equation}
It should be definitely interesting to test this experimentally. 

The left lower panel shows the corresponding situation but for a nonlinearity value less than that required for the coincidence of the stationary state and the initial state. Here we see the selftrapping transition (solid line) sandwiched by the oscillations on both sides for higher and lower $\chi$ values. Notice that both panels exhibit noteworthy differences relative to the degenerate dimer predictions in the shapes of the oscillations. The uppermost (dotted line) curve in the right panel exhibits the `repulsion effect'
\cite{TsiKen1988,*Ken1993}. This means that the initially more populated trap tends to get more (rather than less) populated relative to the less populated one. The origin of this behavior can be understood directly from an explanation given in \cite{TsiKen1988} in terms of what happens in a linear nonresonant system and in \cite{rkb1997} graphically in terms of motion in a potential. 


\section{\label{sec:summary} Remarks}
There are a number of interesting theoretical results that we have obtained in addition. 
\begin{figure}[h!]
\begin{center}
\includegraphics[width=0.8\textwidth]{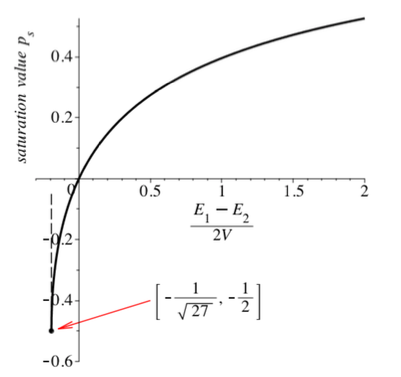}
\caption{The saturation value of the probability difference at the selftrapping transition, $p_s$, shown as tending, with infinite slope, to a calculable value, here $-0.5$, at the critical point as the static energy mismatch, $\Delta/2V$, is varied. The plot is shown for illustrative purposes. See text.}
\label{fig:ps}
\end{center}
\end{figure}
An example is of the value (its dependence on the initial conditions) of $p_s$ at which the probability difference saturates at long times at the selftrapping transition and the striking manner it tends to that value (its dependence on the static energy mismatch with \textit{infinite} slope) at the critical point. Details can be provided in the future if experiments are successfully designed to probe these predictions. For now, as no more than a description of predictions that are available we show in Fig. \ref{fig:ps}  the remarkable prediction for the variation and limit of $p_s$ although we make it here for a localized initial condition.

In summary, we have investigated the tunneling dynamics governing two weakly linked energetically different Bose-Einstein condensates forming a non-degenerate boson Josephson junction. Based on analytical results and explicit plots, we have described novel observable effects. These take the appearance of  \textit{statics} (through the stationary states of the system) as well as \textit{dynamics} through the  dependence of the frequency of quantum oscillations on nonlinearity, the movement of critical points and a critical line of transitions in parameter space, and the time dependence, in some cases peculiar, of the population difference between the two wells (states) of the condensates. These effects show experimentally verifiable predictions of the underlying theoretical concepts. 

\section{\label{sec:disclaimer} Disclaimer}
S. Raghavan contributed to this article in his personal capacity. The views expressed are his own and do not necessarily represent the views of the United States, the Department of Defense (DOD), or its components.

\bibliography{ARKpaper}

\end{document}